\documentclass[12pt]{article}

\usepackage{amsmath,amsfonts,amssymb,latexsym}
\def\mco{\multicolumn}

\numberwithin{equation}{section}
\def\be{\begin{equation}}
\def\ee{\end{equation}}
\def\bq{\begin{eqnarray}}
\def\eq{\end{eqnarray}}
\def\beq{\begin{eqnarray*}}
\def\eeq{\end{eqnarray*}}

\def\a{\alpha}
\def\b{\beta}

\def\d{\delta}

\begin{document}
\begin{titlepage}
\begin{center}

{\Huge Brane singularities and their avoidance}

\vspace{0.8cm}

{\large Ignatios Antoniadis$^{1,*,3}$, Spiros Cotsakis$^{2,\dagger}$, Ifigeneia Klaoudatou$^{2,\ddagger}$}\\

\vspace{0.5cm}

$^1$ {\normalsize {\em Department of Physics, CERN - Theory Division}}\\
{\normalsize {\em CH--1211 Geneva 23, Switzerland}}\\

\vspace{2mm}

$^2$ {\normalsize {\em Research Group of Geometry, Dynamical Systems
and Cosmology}}\\ {\normalsize {\em Department of Information and
Communication Systems Engineering}}\\ {\normalsize {\em University
of the Aegean}}\\ {\normalsize {\em Karlovassi 83 200, Samos,
Greece}}\\
\vspace{2mm} {\normalsize {\em E-mails:}
$^*$\texttt{ignatios.antoniadis@cern.ch},
$^\dagger$\texttt{skot@aegean.gr},
$^\ddagger$\texttt{iklaoud@aegean.gr}}
\end{center}
\begin{abstract}
\noindent The singularity structure and the corresponding asymptotic
behavior of a 3-brane coupled to a scalar field or to a perfect
fluid in a five-dimensional bulk is analyzed in full generality
using the method of asymptotic splittings. In the case of the scalar
field, it is shown that the collapse singularity at a finite
distance from the brane can be avoided only at the expense of making
the brane world-volume positively or negatively curved. In the case
where the bulk field content is parametrized by an analogue of
perfect fluid with an arbitrary equation of state $P=\gamma\rho$
between the `pressure' $P$ and the `density' $\rho$, our results
depend crucially on the constant fluid parameter $\gamma$: (i) For
$\gamma>-1/2$, the flat brane solution suffers from a collapse
singularity at finite distance, that disappears in the curved case.
(ii) For $\gamma<-1$, the singularity cannot be avoided and it
becomes of the big rip type for a flat brane. (iii) For
$-1<\gamma\le -1/2$, the surprising result is found that while the
curved brane solution is singular, the flat brane is not, opening
the possibility for a revival of the self-tuning proposal.

\end{abstract}


$^3${\small On leave from {\em CPHT (UMR CNRS 7644) Ecole
Polytechnique, F-91128 Palaiseau}}

\end{titlepage}

\section{Introduction}
Some time ago, an interesting idea to address the cosmological
constant problem was proposed, based on the so-called self-tuning
mechanism~\cite{nima,silver}. The simplest model consists of a
3-brane embedded in a five-dimensional bulk, in the presence of a
scalar field. The latter is coupled to the brane in a particular
way, motivated by string theory, that allows a flat brane
world-volume solution independently of the brane tension value. It
was, however, realized that a singularity appears in the bulk at
some finite distance from the brane, which can also be thought as a
reservoir through which the vacuum energy decays.

An obvious question is then whether the development of such a
singularity is a generic feature of these models, or under what
conditions may be avoided. In this paper, we investigate this
question in a generalized class of models. Since in this case a
general solution cannot be found analytically, we use a powerful
tool developed a few years ago, called the method of asymptotic
splittings, that allows one to compute all possible asymptotic
behaviors of solutions to the equations of motion around the assumed
location of a singularity~\cite{skot}.

As a first step in our analysis, we consider an extended version of
the simplest model allowing for curved brane world-volume. We then
show that the emergence of the finite-distance singularity is the
\emph{only} possible asymptotic behavior for a flat brane, whereas
for a curved brane the singularity is shifted to an infinite
distance. We also provide a detailed study of the asymptotics of
this model using the method of asymptotic splittings expounded
in~\cite{skot}. A preliminary version of these results was published
in~\cite{iklaoud_6}.

Next, we extend the previous analysis to the case where the bulk
matter is described by an analogue of a perfect fluid with an
arbitrary equation of state $P= \gamma\rho$. In fact, the case of a
massless bulk scalar is a  particular case of such a fluid,
corresponding to the value $\gamma=1$. Here, on the  one hand, we
are interested in the general dynamics of `evolution' of such a
brane-world analogue to cosmology for arbitrary $\gamma$, in order
to reveal the various types of singularity that may develop within a
finite distance from the original position of the brane, and on the
other hand we seek to determine conditions that may lead to the
avoidance of the singularities shifting them to infinite distance
away from the brane.

In particular, we shall show that the existence of a perfect fluid
in the bulk enhances the dynamical possibilities of brane evolution
in the fluid bulk. Such possibilities stem from the different
possible behaviors of the fluid density and the derivative of the
warp factor with respect to the extra dimension. The result depends
crucially on the values of the parameter $\gamma$.

In general, we find three regions of $\gamma$ leading to qualitatively different results:
\begin{itemize}
\item In the region $\gamma>-1/2$, the situation is very similar to the case of a massless bulk scalar field.
Here, the main result we establish is twofold:
\subitem{\bf -} The existence of a singularity at a finite distance is unavoidable in all
solutions with a flat brane. This confirms and extends the results of earlier
works that made similar investigations in different models, using
other methods~\cite{Gubser,Forste}.
\subitem{\bf -} The singularity can be avoided ({\em e.g.} moved at infinite distance)
when the brane becomes curved, either positively or negatively.
Thus, requiring absence of singularity brings back the cosmological constant problem,
since the brane curvature depends on its tension that receives quartically divergent quantum corrections.
\item The situation changes drastically in the region $-1<\gamma<-1/2$. The curved brane solution becomes singular while the flat brane is regular. Thus, this region seems to avoid the main obstruction of the self-tuning proposal: any value of the brane tension is absorbed in the solution and the brane remains flat. The main question is then whether there is a field theory realization of such a fluid producing naturally an effective equation of state of this type.
\item Finally, in the region $\gamma<-1$, corresponding to  the analog of a phantom equation of state,
we show that it is possible for the brane to be ripped apart in as
much the same way as in a big  rip singularity. This happens only in
the flat case, while curved brane solutions develop  `standard'
collapse singularities. No regular solution is found in this region.
\end{itemize}
Besides the above regions, the values $\gamma=-1/2,-1$ are of
special significance: when $\gamma=-1/2$,  we again find a regular
flat brane solution with the so-called sudden behavior~\cite{ba04},
as well as a non-singular curved brane, while for $\gamma=-1$ there
is only singular curved solution.

As mentioned above, it would be very interesting to understand
whether there are field theory representations reproducing the
`exotic' regions of $\gamma\le -1/2$. Obviously the analogy of the
perfect fluid concerning the positivity energy conditions does not
seem to apply in this case where time is replaced by an additional
space coordinate. However, some restrictions may be applied from
usual field theory axioms. Also, the formation of singularities
discussed here is better understood in a dynamical rather than the
usual geometric sense met in general relativity. In the latter case,
cosmological singularities are forming together with conjugate (or
focal) points in spacetime, and for this it is necessary that there
exists at least one timelike dimension (and any number of spacelike
ones, greater than two). The timelike dimension forces then the
geodesics to focus along it rather than along any of the spacelike
dimensions. In the problem discussed in this paper, the timelike
dimension is on the brane while the singularities are forming along
spacelike dimensions in the bulk. As we show below these
singularities are real in the dynamical sense that some component of
the solution vector $(a, a', \rho)$ diverges there. Therefore, we
abandon the usual interpretation according to which the universe
comes to an end in a finite time possibly  through geodesic
refocusing, and instead we study how dynamical effects guide our
brane systems to extreme behaviors.

The structure of this paper is as follows: In Section 2, we analyze
the case of a bulk scalar field. We first choose appropriate
variables and rewrite the basic field equations in the form of a
dynamical system; secondly, we introduce some convenient terminology
for the different types of singularity to be met later in our
analysis. Then, in the Subsections 2.1 and 2.2, we give the
asymptotics of the models consisting of flat and curved brane
respectively. In Section 3, we study the case in which the bulk is
filled with an analogue of a perfect fluid. We first derive the form
of the dynamical system and single out the possible dominant
balances, organizing centers of all the evolutionary behaviors that
fully characterize this case. In Subsections 3.1 and 3.2, we study
carefully the asymptotics around collapse singularities of two
types, that we call type I and type II singularities, respectively.
In Subsection 3.3, we explore the dynamics as the brane approaches a
big rip singularity, while in Subsection 3.4 we look at a milder
singularity that resembles in many ways the so-called sudden
(non-singular) behavior introduced in Ref.~\cite{ba04}. In
Subsection 3.5, we analyze the possibility of avoiding
finite-distance singularities leading to the existence of regular
brane evolution in the bulk. Finally, in Section 4 we conclude and
also comment on possible future work in various directions,
considering for instance other forms of matter in the bulk. In an
Appendix, we briefly outline the basic steps of the method of
asymptotic splittings.
\section{Dynamics of scalar field-brane configuration}
In this Section, we study dynamical aspects of a braneworld model
$(\mathcal{V}_4\equiv\mathbb{R}\times\mathcal{M}_3,g_4)$ consisting
of a 3-brane $(\mathcal{M}_3,g_3)$ embedded in a five-dimensional
bulk space $(\mathcal{V}_5,g_5)$ with a scalar field minimally
coupled to the bulk.  The total action $S_{total}$ splits in two
parts, namely, the bulk action $S_{bulk}$ and the brane action
$S_{brane}$, \be \label{s_tot} S_{total}=S_{bulk}+S_{brane}, \ee
with \bq S_{bulk}&=&\int \left(\frac{R}{2\kappa^{2}_{5}}-
\frac{\lambda}{2}(\nabla\phi)^{2}\right)d\mu_{g_{5}},\\
S_{brane}&=&-\int f(\phi)d\mu_{g_{4}},\,\,
\textrm{at}\,\,Y=Y_{\ast}, \eq
where the measures $d\mu_{g_{5}}=d^{4}x
dY\sqrt{\textrm{det}g_{5}}$, and
$d\mu_{g_{4}}=d^{4}x\sqrt{\textrm{det}g_{4}}$,  $Y$ denotes the fifth bulk dimension, $Y_{\ast}$ is
the assumed initial position of the brane, $\lambda$ is a parameter
defining the type of scalar field $\phi$,
$\kappa^{2}_{5}=M_{5}^{-3}$, $M_{5}$ being the five-dimensional
Planck mass, and $f(\phi)$ denotes the tension of the brane as a
function of the scalar field.

Varying the total action (\ref{s_tot}) with respect to $g_{5}$, we
find the five-dimensional Einstein field equations in the form
~\cite{iklaoud_6}, \be \label{einst5d}
R_{AB}-\frac{1}{2}g_{AB}R=\lambda \kappa_{5}^{2} \left(\nabla
_{A}\phi\nabla_{B}\phi-\frac{1}{2}g_{AB}(\nabla\phi)^{2}\right)
+\frac{2\kappa_{5}^{2}}{\sqrt{\textrm{det}g_{5}}} \frac{\delta
(\sqrt{\textrm{det}g_{4}}f(\phi))} {\delta
g^{\a\b}}\,\delta_{A}^{\a}\delta_{B}^{\b}\delta(Y), \ee while the
scalar field equation is obtained by variation of the action
(\ref{s_tot}) with respect to $\phi$~\cite{iklaoud_6} and it is: \be
\label{scalarbr} \lambda \Box
_{5}\phi=-\frac{1}{\sqrt{\textrm{det}g_{5}}}\frac{\delta
(\sqrt{\textrm{det}g_{4}}f(\phi))} {\delta\phi}\,\delta(Y), \ee where
$A,B=1,2,3,4,5$ and $\a,\b=1,2,3,4$ while $\delta (Y)=1$ at
$Y=Y_{\ast}$ and vanishing everywhere else, and \be \Box _{5}\phi=
\frac{1}{\sqrt{\textrm{det}g_{5}}}\nabla_{A}(\sqrt{\textrm{det}g_{5}}
g^{AB}\nabla_{B}\phi) .
\ee

In the following, we assume a bulk metric of the form
\be
\label{warpmetric}
g_{5}=a^{2}(Y)g_{4}+dY^{2},
\ee
where $g_{4}$ is the four-dimensional flat, de Sitter or anti de Sitter metric,
{\it i.e.},
\be
\label{branemetrics}
g_{4}=-dt^{2}+f^{2}_{k}g_{3},
\ee
with
\be
\label{g_3}
g_{3}=dr^{2}+h^{2}_{k}g_{2},
\ee
and
\be
\label{g_2}
g_{2}=d\theta^{2}+\sin^{2}\theta d\varphi^{2}.
\ee
Here $f_{k}=1,\cosh (H t)/H,\cos (H t)/H $ ($H^{-1}$ is the de Sitter
curvature radius) and $ h_{k}=r,\sin r,\sinh r $, respectively.

The field equations (\ref{einst5d}) and (\ref{scalarbr}) then take the
form \bq \label{feq1}
\frac{a'^{2}}{a^{2}}&=&\frac{\lambda\kappa^{2}_{5}\phi'^{2}}{12}+\frac{k
H^{2}}{a^{2}}, \\
\label{feq2}
\frac{a''}{a}&=&-\frac{\lambda\kappa^{2}_{5}\phi'^{2}}{4}, \\
\label{feq3} \phi''+4\frac{a'}{a}\phi'&=&0, \eq where the prime $(')$
denotes differentiation with respect to $Y$, and $k=0,\pm 1$. The
variables to be determined are $a$, $a'$ and $\phi'$. These three
equations are not independent since Eq.~(\ref{feq2}) was derived
after substitution of Eq.~(\ref{feq1}) in the field equation
$G_{\a\a}=\kappa_{5}^{2}T_{\a\a}$, $\a=1,2,3,4$, \be
\frac{a''}{a}+\frac{a'^{2}}{a^{2}}-\frac{kH^{2}}{a^{2}}
=-\lambda\kappa_{5}^{2}\frac{\phi'^{2}}{6}. \ee
In our analysis below we use the independent equations (\ref{feq2})
and (\ref{feq3}) to determine the unknown variables $a$, $a'$ and
$\phi'$, while Eq.~(\ref{feq1}) will then play the role of a
constraint equation for our system.

Assuming a $Y\rightarrow -Y$ symmetry and solving the Eqs.~(\ref{einst5d})
(the -$\a\a$- component, $\a=1,2,3,4$) and
(\ref{scalarbr}) on the brane we get \bq \label{bound1}
a'(Y_{\ast})&=&-\frac{\kappa_{5}^{2}}{6}f(\phi(Y_{\ast}))a(Y_{\ast}),\\
\label{bound2}
\phi'(Y_{\ast})&=&\frac{f'(\phi(Y_{\ast}))}{2\lambda}. \eq The
particular coupling used in~\cite{nima} allows only for flat
solutions to exist.
This easily follows by using equations (\ref{bound1}) and
(\ref{bound2}) and solving the FRW equation (\ref{feq1}) on the
brane for $kH^{2}$:
$$kH^{2}=\frac{a^{2}(Y_{\ast})\kappa^{2}_{5}}{12}
\left(\frac{\kappa_{5}^{2}}{3}f^{2}(\phi(Y_{\ast}))-
\frac{f'^{2}(\phi(Y_{\ast}))}{4\lambda}\right).$$ Clearly, $k$ is
identically zero if and only if:
$$\frac{f'(\phi)}{f(\phi)}=2\sqrt{\frac{\lambda}{3}}\kappa_{5},$$ or equivalently, if and
only if $f(\phi)\propto e^{2\sqrt{\lambda/3}\kappa_{5}\phi}$ (the
authors of~\cite{nima} have set $\lambda=3$ and hence the
appropriate choice for the brane tension in that case is
$f(\phi)\propto e^{2\kappa_{5}\phi}$). In our more general problem,
the coupling function cannot be fixed this way. By working with
other couplings we can allow for non-flat, maximally symmetric
solutions to exist and avoid having the singularity at a finite
distance away from the position of the brane.

Our purpose is to find all possible asymptotic behaviors around the assumed
position of a singularity, denoted by $Y_{s}$, emerging from general or
particular solutions of the system (\ref{feq1})-(\ref{feq3}). The most useful
tool for this analysis is the method of asymptotic splittings~\cite{skot}
(see the Appendix for a brief introduction), in which we start by setting \be
x=a, \quad y=a', \quad z=\phi'. \ee The field equations (\ref{feq2})
and (\ref{feq3}) become the following system of ordinary
differential equations: \bq \label{syst1_1}
x'&=&y \\
\label{syst1_2}
y'&=&-\lambda Az^{2}x \\
\label{syst1_3} z'&=&-4y\frac{z}{x}, \eq where $A=\kappa^{2}_{5}/4$.
Hence, we have a dynamical system determined by the non-polynomial
vector field \be \mathbf{f}=\left(y,-\lambda
Az^{2}x,-4y\frac{z}{x}\right)^{\top}. \ee Equation (\ref{feq1}) does
not include any terms containing derivatives with respect to $Y$; it
is a constraint equation which in terms of the new variables takes
the form \be \label{constraint1}
\frac{y^{2}}{x^{2}}=\frac{A\lambda}{3} z^{2}+\frac{k H^{2}}{x^{2}}.
\ee Equations (\ref{syst1_1})-(\ref{syst1_3}) and
(\ref{constraint1}) constitute the basic dynamical system of our
study in this Section.

Before we proceed with the analysis of the above system,
we introduce the following terminology, which we use in subsequent paragraphs, for
the possible singularities to occur at a finite-distance from the brane. Specifically, we call
a state where:
\begin{enumerate}
\item[i)]$  $ $a\rightarrow 0$, $a'\rightarrow \infty$ and
$\rho\rightarrow \infty$: a singularity of collapse type I.
\item[ii)]$  $ $a\rightarrow 0$, $a'\rightarrow a'_{s}$ and
$\rho\rightarrow 0$: a singularity of collapse type IIa,\\
\indent$a\rightarrow 0$, $a'\rightarrow a'_{s}$ and
$\rho\rightarrow \rho_{s}$: a singularity of collapse type IIb,\\
\indent$a\rightarrow 0$, $a'\rightarrow a'_{s}$ and
$\rho\rightarrow \infty$: a singularity of collapse type IIc,\\
where $a'_{s}$, $\rho_{s}$ are non-vanishing finite constants.
\item[iii)]$  $ $a\rightarrow \infty$, $a'\rightarrow -\infty$ and
$\rho\rightarrow\infty$: a big rip singularity,
\end{enumerate}
where $\rho$ denotes the density of the matter component in the bulk that is
considered each time. For the case of interest in this Section, $\rho=\lambda\phi'^2/2$ and
we shall show that there are two major cases to be treated, the first is when we
choose $k=0$ in (\ref{constraint1}) and corresponds to a brane being
flat, while in the second case $k\neq 0$, giving constant curvature
to the brane. We shall treat these two cases independently. One important result
of our analysis of this system will be
that the inclusion of nonzero curvature for the brane moves the
singularity (that is of the collapse type I class) an infinite distance away
from the brane.
\subsection{Collapse type I singularity: Flat brane}
In this Subsection we take $k=0$ in the constraint equation (\ref{constraint1}),
\be
\label{constraint_flat} \frac{y^{2}}{x^{2}}=\frac{A\lambda}{3}
z^{2}. \ee We shall show that the only possible asymptotic behavior
of the solutions of this system (flat brane) is that $a\rightarrow
0$, $a'\rightarrow \infty$ and $\phi'\rightarrow \infty$, as
$Y\rightarrow Y_{s}$.

Following the basic steps of the method of asymptotic splittings expounded in the Appendix,
we start our asymptotic analysis by inserting in the system
(\ref{syst1_1})-(\ref{syst1_3}) the forms
\be
\label{dominant forms} (x,y,z)=(\alpha\Upsilon^{p},\beta
\Upsilon^{q},\zeta\Upsilon^{m}),
\ee
where $\Upsilon=Y-Y_{s}$ and
\be
\label{pqm}
(p,q,m)\in\mathbb{Q}^{3}
\quad \textrm{and} \quad (\alpha,\beta,\zeta)\in
\mathbb{C}^{3}\smallsetminus\{\mathbf{0}\}.
\ee
We find that in the neighborhood of the singularity the only possible dominant
balance, that is pairs of the form
\be
\label{pairs}
\mathcal{B}=\{\mathbf{a},\mathbf{p}\}, \quad \textrm{where}
\quad \mathbf{a}=(\alpha,\beta,\delta), \quad \mathbf{p}=(p,q,r),
\ee
determining the dominant asymptotics as we approach the singularity,
is the following: \be \label{sing}
\mathcal{B}_{1}=\{(\alpha,\alpha/4,\sqrt{3}/(4\sqrt{A\lambda})),
(1/4,-3/4,-1)\}. \ee Note that a second balance $\mathcal{B}_{2}$
becomes only possible when we allow for nonzero curvature, $k\neq
0$, and will be analyzed in the next Subsection. There are no other
acceptable balances, hence all the possible asymptotic behaviors for
a flat and curved brane can be described uniquely by the balances
$\mathcal{B}_{1}$ and $\mathcal{B}_{2}$ respectively. Our purpose is
then to construct asymptotic expansions of solutions
in the form of a series defined by
\be \label{Puiseux}
\mathbf{x}=\Upsilon^{\mathbf{p}}(\mathbf{a}+
\Sigma_{j=1}^{\infty}\mathbf{c}_{j}\Upsilon^{j/s}),
\ee
where $\mathbf{x}=(x,y,z)$, $\mathbf{c}_{j}=(c_{j1},c_{j2},c_{j3})$, and
$s$ is 
the least common multiple of the denominators of
the positive $\mathcal{K}$-exponents (cf. \cite{skot},
\cite{goriely}).

First we calculate the Kowalevskaya matrix ($\mathcal{K}$-matrix in short),
given by
\be
\mathcal{K}=D\mathbf{f}(\mathbf{a})-\textrm{diag}(\mathbf{p}),
\ee
where $D\mathbf{f}(\mathbf{a})$ is the Jacobian matrix of
$\mathbf{f}$, which in our case reads:
\be
D\mathbf{f}(x,y,z)=\left(
                     \begin{array}{ccc}
                       0 & 1             & 0 \\
               -A\lambda z^{2} & 0             & -2A\lambda x z \\
       \dfrac{4y z}{x^{2}} & -\dfrac{4z}{x} & -\dfrac{4y}{x} \\
                     \end{array}
                   \right),
\ee
to be evaluated on $\mathbf{a}$. For the $\mathcal{B}_{1}$
balance we have that
$\mathbf{a}=(\alpha,\alpha/4,\sqrt{3}/(4\sqrt{A\lambda}))$, and
$\mathbf{p}=(1/4,-3/4,-1)$, thus giving
\be
\mathcal{K}_{1}=\left(
                     \begin{array}{ccc}
                       -\dfrac{1}{4} & 1                                   & 0 \\
                      -\dfrac{3}{16} & \dfrac{3}{4}             & -\dfrac{\sqrt{3A\lambda}\a}{2} \\
\dfrac{\sqrt{3}}{4\a\sqrt{A\lambda}} & -\dfrac{\sqrt{3}}{\a\sqrt{A\lambda}} & 0 \\
                     \end{array}
                   \right).
\ee

The next step is to calculate the $\mathcal{K}$-exponents for this balance.
These exponents are the eigenvalues of the $\mathcal{K}_{1}$ matrix and
constitute its spectrum, $spec(\mathcal{K}_{1})$.
The arbitrary constants of any (particular or general) solution first appear in
those terms in the series (\ref{Puiseux}) whose coefficients $\mathbf{c}_{k}$ have indices $k=\varrho s$,
where $\varrho$ is a non-negative $\mathcal{K}$-exponent. The number of
non-negative $\mathcal{K}$-exponents equals therefore
the number of arbitrary constants that appear in the series expansions
(\ref{Puiseux}). There is always the $-1$ exponent that corresponds to an
arbitrary constant that is the position of the singularity, $Y_{s}$. The
balance $\mathcal{B}_{1}$ corresponds thus to a general solution in our
case if and only if it possesses two non-negative $\mathcal{K}$-exponents
(the third arbitrary constant is the position of the singularity, $Y_{s}$).
Here we find
\bq
\textrm{spec}(\mathcal{K}_{1})=\{-1,0,3/2\}
\eq
so that $\mathcal{B}_{1}$ indeed corresponds to a general solution.
After substituting in the system (\ref{syst1_1})-(\ref{syst1_3}) the series
expansions
\be
x=\Sigma_{j=0}^{\infty}\,c_{j1}\Upsilon^{j/2+1/4},\quad
y=\Sigma_{j=0}^{\infty}\,c_{j2}\Upsilon^{j/2-3/4},\quad
z=\Sigma_{j=0}^{\infty}\,c_{j3}\Upsilon^{j/2-1},
\ee
we arrive at the following asymptotic solution around the singularity:
\bq \label{Puis_1x}
x&=&\alpha\Upsilon^{1/4}+\frac{4}{7}c_{32}\Upsilon^{7/4}+\cdots \\
y&=&\frac{\alpha}{4}\Upsilon^{-3/4}+c_{32}\Upsilon^{3/4}+\cdots\\
\label{Puis_1z} z&=&\frac{\sqrt{3}}{4\sqrt{A\lambda}}\Upsilon^{-1}-
\frac{4\sqrt{3}}{7\alpha\sqrt{A\lambda}}c_{32}\Upsilon^{1/2}+\cdots.
\eq

The last step is to check whether for each $j$ satisfying
$j/2=\varrho$ with $\varrho$ a positive eigenvalue, the
corresponding eigenvector $v$ of the $\mathcal{K}_{1}$ matrix is
such that the compatibility conditions hold, namely, \be
v^{\top}\cdot P_{j}=0, \ee where $P_{j}$ are polynomials in
$\mathbf{c}_{1},\ldots, \mathbf{c}_{j-1}$ given by \be
\mathcal{K}_{1}\mathbf{c}_{j}-(j/s)\mathbf{c}_{j}=P_{j}. \ee Here
the relation $j/2=3/2$ is valid only for $j=3$ and the associated
eigenvector is \be v^{\top}=\left(-\frac{\a\sqrt{A\lambda}}{\sqrt{3}},
-\frac{7\a\sqrt{A\lambda}}{4\sqrt{3}},1\right). \ee The
compatibility condition \be v^{\top}\cdot
(\mathcal{K}_{1}-(3/2)\mathcal{I}_{3})\mathbf{c}_{3}=0, \ee
therefore indeed holds since \be
(\mathcal{K}_{1}-(3/2)\mathcal{I}_{3})\mathbf{c}_{3}=c_{32} \left(
  \begin{array}{ccc}
    -\dfrac{7}{4} & 1 & 0 \\ \\
    -\dfrac{3}{16} & -\dfrac{3}{4} & -\dfrac{\a\sqrt{3A\lambda}}{2} \\ \\
    \dfrac{\sqrt{3}}{4\a\sqrt{A\lambda}} & -\dfrac{\sqrt{3}}{\a\sqrt{A\lambda}} & -\dfrac{3}{2} \\
  \end{array}
\right) \left(
  \begin{array}{c}
    \dfrac{4}{7} \\ \\
    1 \\ \\
     -\dfrac{4\sqrt{3}}{7\a\sqrt{A\lambda}}\\
  \end{array}
\right)=\left(
          \begin{array}{c}
            0 \\ \\
            0 \\ \\
            0 \\
          \end{array}
        \right).
\ee This shows that a representation of the solution asymptotically
by a Puiseux series as given in eqs.~(\ref{Puis_1x})-(\ref{Puis_1z})
is valid. Hence, we conclude that at finite distance from the brane, a
collapse type I singularity develops, {\it i.e.}, as $Y\rightarrow Y_{s}$ the
asymptotic forms of the variables are:
\be
\label{behscI} a\rightarrow 0, \quad
a'\rightarrow\infty, \quad \phi'\rightarrow \infty.
\ee
This is exactly the asymptotic behavior of the solution found previously by
Arkani-Hamed \emph{et al} in~\cite{nima}. Our analysis shows that
this is \emph{the only possible} asymptotic behavior for a flat
brane since there exist no other dominant balances in this case.
\subsection{Behavior at infinity: Curved brane}
In this Subsection we show that the collapse type I singularity that
necessarily arises in the case of a flat brane is avoided (or
shifted at an infinite distance away from the brane) when we
consider a curved brane instead.

The new asymptotics follow from the study of a second balance that
results from the substitution of (\ref{dominant forms}) in
(\ref{syst1_1})-(\ref{syst1_3}). We calculate this new balance to
be \be \label{nonsing}
\mathcal{B}_{2}=\{(\alpha,\alpha,0),(1,0,-1)\}. \ee It corresponds
to a particular solution for a \emph{curved brane} since it
satisfies Eq.~(\ref{constraint1}) for $k\neq 0$ and $\alpha^{2}=k
H^{2}$ (here we have to sacrifice one arbitrary constant by setting
it equal to $kH^{2}$), $k=\pm 1$. The $\mathcal{K}$-matrix of
$\mathcal{B}_{2}$ is \be
\mathcal{K}_{2}=D\mathbf{f}(\a,\a,0)-\textrm{diag}(1,0,-1)=\left(
  \begin{array}{ccc}
    -1 & 1 &  0 \\
    0 & 0 &  0 \\
    0 & 0 & -3 \\
  \end{array}
\right), \ee with eigenvalues \be
\textrm{spec}(\mathcal{K}_{2})=\{-1,0,-3\}. \ee
Thus for the balance $\mathcal{B}_{2}$ we find two distinct, negative, integer
$\mathcal{K}$-exponents and an infinite expansion in negative powers
of a \emph{particular} solution (recall that we had to sacrifice one
arbitrary constant) around the presumed singularity at $Y_{s}$,
with the negative $\mathcal{K}$-exponents signaling the positions
where the arbitrary constants first appear~\cite{fordy}. We
therefore expand the variables in series with descending powers of
$\Upsilon$ in order to meet the two arbitrary constants occurring
for $j=-1$ and $j=-3$, {\it i.e.}, \be
x=\Sigma_{j=0}^{-\infty}c_{j1}\Upsilon^{j+1}, \quad
y=\Sigma_{j=0}^{-\infty}c_{j2}\Upsilon^{j}, \quad
z=\Sigma_{j=0}^{-\infty}c_{j3}\Upsilon^{j-1}. \ee Substituting these
series expansions back in the system (\ref{syst1_1})-(\ref{syst1_3})
and after some manipulation, we find the following asymptotic
behavior \bq \label{Puis_2x}
x&=&\alpha\Upsilon+c_{-1\,1}+\cdots \\
y&=&\alpha+\cdots \\
\label{Puis_2z} z&=&c_{-3\,3}\Upsilon^{-4}+\cdots . \eq It is easy to check
the compatibility conditions for $j=-1$ and $j=-3$.
We find that \be (\mathcal{K}_{2}+\mathcal{I}_{3})\mathbf{c}_{-1}=\left(
                                                \begin{array}{ccc}
                                                  0 & 1 &  0 \\
                                                  0 & 1 &  0 \\
                                                  0 & 0 & -2 \\
                                                \end{array}
                                              \right)
\left(
  \begin{array}{c}
    c_{-11} \\
    0 \\
    0 \\
  \end{array}
\right)=\left(
          \begin{array}{c}
            0 \\
            0 \\
            0 \\
          \end{array}
        \right),
\ee and \be (\mathcal{K}_{2}+3\mathcal{I}_{3})\mathbf{c}_{-3}=\left(
                                                \begin{array}{ccc}
                                                  2 & 1 & 0 \\
                                                  0 & 3 & 0 \\
                                                  0 & 0 & 0 \\
                                                \end{array}
                                              \right)
\left(
  \begin{array}{c}
    0 \\
    0 \\
    c_{33} \\
  \end{array}
\right)=\left(
          \begin{array}{c}
            0 \\
            0 \\
            0 \\
          \end{array}
        \right),
\ee so that the compatibility conditions are indeed satisfied. The
expansions given by Eqs.~(\ref{Puis_2x})-(\ref{Puis_2z}) are
therefore valid, and we can say that as $\Upsilon\rightarrow 0$, or
equivalently as $S\equiv 1/\Upsilon\rightarrow \infty$, we have that
\be \label{behscII} a\rightarrow \infty, \quad a'\rightarrow \infty,
\quad \phi'\rightarrow \infty. \ee Therefore for a curved brane we
find that there can be no finite-distance singularities. The only
possible asymptotic behavior is the one given in (\ref{behscII})
which is only valid at an infinite distance from the brane.

It is interesting that apart from the balances (\ref{sing}) and (\ref{nonsing}),
a third balance which initially arises from
the substitution of (\ref{dominant forms}) in
(\ref{syst1_1})-(\ref{syst1_3}), namely, the form
$$\mathcal{B}_{3}=\{(\alpha,0,0),(0,-1,-1)\},$$ is not acceptable
in this case of bulk scalar field, since it does not give the
necessary $-1$ $\mathcal{K}$-exponent.
If this balance were acceptable it would yield a finite-distance behavior of the type $a\rightarrow \alpha$, $a'\rightarrow a'_{s}$, $\phi'\rightarrow \phi'_{s}$, where $\alpha$ is the constant appearing in the balance $\mathcal{B}_{3}$ and $a'_{s}$, $\phi'_{s}$ are also constants. This would be 
similar to the sudden behavior met in classical four-dimensional cosmologies with a perfect fluid \cite{ba04}, in the sense that the warp factor, in analogy with the scale factor, its derivative and the density all remain finite.
Below, we find that such a balance 
\emph{does} become possible 
when we replace the scalar field with a perfect fluid.

\section{Dynamics in a perfect fluid bulk}
In this Section, we rewrite the brane model living in a bulk, which we now consider
to be filled with a perfect fluid, as a dynamical system in three basic
variables and completely identify the principal modes of approach to its
singularities, that is we find all the dominant balances of the
system. In this case, bulk space is filled with a perfect fluid with equation of state
$P=\gamma \rho$, where the pressure $P$ and the density $\rho$ are
functions only of the fifth dimension, $Y$. We assume again a
bulk metric of the form (\ref{warpmetric})-(\ref{g_2})
and an energy-momentum tensor of the form
$T_{AB}=(\rho+P)u_{A}u_{B}-Pg_{AB}$, where $A,B=1,2,3,4,5$ and
$u_{A}=(0,0,0,0,1)$, with the 5th coordinate corresponding to $Y$.

The five-dimensional Einstein equations, \be
G_{AB}=\kappa^{2}_{5}T_{AB}, \ee
can be written in the
following form: \bq \label{syst2i}
\frac{a''}{a}&=&-\kappa^{2}_{5}\frac{(1+2\gamma)}{6}\rho, \\
\label{syst2iii} \frac{a'^{2}}{a^{2}}&=&\frac{\kappa^{2}_{5}}{6}
\rho+\frac{k H^{2}}{a^{2}}, \eq
where as in the case treated previously $k=0,\pm 1$, and the prime
$(\,')$ denotes differentiation with respect to $Y$. The equation of
conservation, \be \nabla_{B}T^{AB}=0, \ee becomes \be
\label{syst2ii} \rho'+4(1+\gamma)\frac{a'}{a}\rho=0. \ee Introducing the new
variables \be x=a, \quad y=a', \quad w=\rho, \ee Eqs.~(\ref{syst2i})
and (\ref{syst2ii}) take the form \bq \label{syst2a}
x'&=&y, \\
y'&=&-2A\frac{(1+2\gamma)}{3}w x, \\
\label{syst2c} w'&=&-4(1+\gamma)\frac{y}{x}w, \eq while
Eq.~(\ref{syst2iii}) reads \be \label{constraint3}
\frac{y^{2}}{x^{2}}=\frac{2}{3} A w+\frac{k H^{2}}{x^{2}}, \quad
A=\kappa_{5}^{2}/4. \ee Since this last equation does not contain
derivatives with respect to $Y$, it is a velocity independent
constraint equation for the system (\ref{syst2a})-(\ref{syst2c}).

The next step is to apply the method of asymptotic splittings in an
effort to find all possible asymptotic behaviors of the dynamical
system (\ref{syst2a})-(\ref{syst2c}) with the constraint
(\ref{constraint3}), by building series expansions of the solutions
around the presumed position of the singularity at $Y_{s}$.
We note that the system (\ref{syst2a})-(\ref{syst2c}) is a weight
homogeneous system determined by the vector field \be
\label{vectorfield} \mathbf{f}=\left(y,-2A\frac{(1+2\gamma)}{3}w
x,-4(1+\gamma) \frac{y}{x}w\right)^{\top}. \ee
In order to compute all possible dominant balances that describe the
principal asymptotics of the system,
\be
\label{dominant forms new} (x,y,w)=(\alpha\Upsilon^{p},\beta
\Upsilon^{q},\zeta\Upsilon^{m}),
\ee
where $\Upsilon=Y-Y_{s}$, we look for pairs of the form
(\ref{pairs}) with (\ref{pqm}),
as we did in Section~2 for the case of the bulk scalar field. 
We find after some calculation the
following list of all possible balances for our basic system
(\ref{syst2a})-(\ref{constraint3}):
\bq
_{\gamma}\mathcal{B}_{1}&=&
\left\{\left(\alpha,\alpha
p,\frac{3}{2A}p^{2}\right),(p,p-1,-2)\right\},
p=\frac{1}{2(\gamma+1)}, \, \gamma \neq -1/2,-1,\label{preveq}\\
_{\gamma}\mathcal{B}_{2}&=&\{(\alpha,\alpha,0),(1,0,-2)\},\quad \gamma \neq -1/2,\\
_{-1/2}\mathcal{B}_{3}&=&\{(\alpha,\alpha,0),(1,0,r)\},
\\
_{-1/2}\mathcal{B}_{4}
&=&\{(\alpha,\alpha,\delta),(1,0,-2)\},\\
_{-1/2}\mathcal{B}_{5}&=&\{(\a,0,0), (0,-1,r)\},\label{Bfive} \eq
where $_{-1/2}\mathcal{B}_{i}\equiv_{\gamma=-1/2}\mathcal{B}_{i}$.
Note that, as already mentioned in the Introduction,
the first balance $_{\gamma}\mathcal{B}_{1}$ for $\gamma=1$
coincides with the balance $\mathcal{B}_{1}$ in eq.~(\ref{sing}),
where the fluid was described by a massless bulk scalar field with an arbitrary
coupling to the brane.

The above balances are \emph{exact} solutions of the system and they
must therefore also satisfy the constraint equation
(\ref{constraint3}). This fact alters the presumed generality of the
solution represented by each one of the balances above and
determines uniquely the type of spatial geometry that we must
consider: The balances $_{\gamma}\mathcal{B}_{1}$ and
$_{-1/2}\mathcal{B}_{5}$ are found when we set $k=0$, and describe a
(potentially general) solution corresponding to a flat brane, while
the balances $_{\gamma}\mathcal{B}_{2}$ and $_{-1/2}\mathcal{B}_{3}$
were found when $k\neq 0$ and describe particular solutions of
curved branes (since we already have to sacrifice the arbitrary
constant $\a$ by imposing $\a^{2}=kH^{2}$). For the balance
$_{-1/2}\mathcal{B}_{4}$, on the other hand, $k$ is not specified
and hence it describes a particular solution for a curved or flat
brane (particularly since we have to set
$\d=(3/(2A))(1-kH^{2}/\a^{2})$ to satisfy eq.~(\ref{constraint3})).

Each one of these balances are analyzed in detail in the following
Subsections according to the nature of asymptotic behaviors they imply.
\subsection{Collapse type I singularity}
We shall focus in this Subsection exclusively on a study of the balance
$_{\gamma}\mathcal{B}_{1}$ and show that for certain ranges of
$\gamma$ it gives the generic asymptotic behavior of a flat brane to
a singularity of collapse type I. Our analysis implies that such
behavior in the case of a fluid bulk can only result from a
$_{\gamma}\mathcal{B}_{1}$ type of balance.

As a first step we calculate the $\mathcal{K}$-matrix,
$\mathcal{K}=D\mathbf{f}(\mathbf{a})-\textrm{diag}(\mathbf{p})$,
where $D\mathbf{f}(\mathbf{a})$ is the Jacobian matrix of
$\mathbf{f}$ \footnote{$\mathbf{f}$ is the vector field resulting
from the dynamical system (\ref{syst2a})-(\ref{syst2c}) and
$\{\mathbf{a},\mathbf{p}\}$ is the balance
$_{\gamma}\mathcal{B}_{1}$.}: We have \be D\mathbf{f}(x,y,w)=\left(
                     \begin{array}{ccc}
                       0 & 1             & 0 \\ \\
               -\dfrac{2}{3}(1+2\gamma)A w & 0             & -\dfrac{2}{3}(1+2\gamma)A x  \\ \\
       4(1+\gamma)\dfrac{y w}{x^{2}} & -4(1+\gamma)\dfrac{w}{x} & -4(1+\gamma)\dfrac{y}{x} \\
                     \end{array}
                   \right),
\ee to be evaluated on $\mathbf{a}$. The balance
$_{\gamma}\mathcal{B}_{1}$ has $\mathbf{a}=(\alpha,\alpha
p,3p^{2}/2A)$, and $\mathbf{p}=(p,p-1,-2)$, with
$p=1/(2(\gamma+1))$. Thus the $\mathcal{K}$-matrix
for this balance is \beq
\quad\quad\quad_{\gamma}\mathcal{K}_{1}&=&D\mathbf{f}\left(\a,\a p,\frac{3}{2A}p^{2}\right)-\textrm{diag}(p,p-1,-2)\\
&=&
D\mathbf{f}\left(a,\frac{a}{2(1+\gamma)},\frac{3}{8A(1+\gamma)^{2}}\right)
-\textrm{diag}\left(\frac{1}{2(1+\gamma)},-\frac{1+2\gamma}{2(1+\gamma)},-2\right)
\eeq \be =\left(
                     \begin{array}{ccc}
                       -\dfrac{1}{2(1+\gamma)} & 1             & 0 \\ \\
               -\dfrac{1+2\gamma}{4(1+\gamma)^{2}} & \dfrac{1+2\gamma}{2(1+\gamma)}           & -\dfrac{2}{3}(1+2\gamma)A \a  \\ \\
       \dfrac{3}{4(1+\gamma)^{2}A \a} & -\dfrac{3}{2(1+\gamma)A \a} & 0 \\
                     \end{array}
                   \right).
\ee

We then calculate what the $\mathcal{K}$-exponents for this
balance actually are. Recall that these exponents are the
eigenvalues of the matrix $_{\gamma}\mathcal{K}_{1}$ and constitute
its spectrum, $spec(_{\gamma}\mathcal{K}_{1})$.
The balance $_{\gamma}\mathcal{B}_{1}$ corresponds to a general
solution in our case if and only if it possesses two non-negative
$\mathcal{K}$-exponents (the third arbitrary constant is the
position of the singularity, $Y_{s}$).
Here we find \be \textrm{spec}(_{\gamma}\mathcal{K}_{1})=
\left\{-1,0,\frac{1+2\gamma}{1+\gamma}\right\}. \ee The last
eigenvalue is a function of the $\gamma$ parameter and it is
positive when either $\gamma<-1$, or $\gamma>-1/2$. We consider here
the case $\gamma>-1/2$ since, as it will soon follow, this range of
$\gamma$ is adequate for the occurrence of a collapse type I
singularity. The case of $\gamma<-1$ leads to a big rip singularity
and will be examined in Subsection 3.3.

Let us assume $\gamma=-1/4$ for concreteness. Then \bq
_{-1/4}\mathcal{B}_{1}&=&\{(\a,2\a/3,2/(3A)),(2/3,-1/3,-2)\},\\
\textrm{spec}(_{-1/4}\mathcal{K}_{1})&=&\{-1,0,2/3\}. \eq
Substituting in the system (\ref{syst2a})-(\ref{syst2c}) the
particular value $\gamma=-1/4$ and the forms \be
x=\Sigma_{j=0}^{\infty}\,c_{j1}\Upsilon^{j/3+2/3}, \quad
y=\Sigma_{j=0}^{\infty}\,c_{j2}\Upsilon^{j/3-1/3}, \quad
w=\Sigma_{j=0}^{\infty}\,c_{j3}\Upsilon^{j/3-2}, \ee we arrive at the
following asymptotic expansions: \bq \label{g_B1x}
x&=& \a\Upsilon^{2/3}-\frac{A\a}{2}c_{2\,3}\Upsilon^{4/3}+\cdots,\\
y&=& \frac{2}{3}\a\Upsilon^{-1/3}-\frac{2}{3}A \a c_{2\,3}\Upsilon^{1/3}+\cdots,\\
\label{g_B1w} w&=&
\frac{2}{3A}\Upsilon^{-2}+c_{2\,3}\Upsilon^{-4/3}+\cdots. \eq For
this to be a valid solution we need to check whether the
compatibility condition holds true for each $j$ satisfying
$j/3=\varrho$ with $\varrho$ a positive eigenvalue. Here the
corresponding relation $j/3=2/3$ is valid only for $j=2$ and the
compatibility condition indeed holds since, \be
(_{-1/4}\mathcal{K}_{1}-(2/3)\mathcal{I}_{3})\mathbf{c}_{2}= \left(
  \begin{array}{ccc}
    -\dfrac{4}{3} & 1 & 0 \\ \\
    -\dfrac{2}{9} & -\dfrac{1}{3} & -\dfrac{A\a}{3} \\ \\
    \dfrac{4}{3A\a} & -\dfrac{2}{A\a} & -\dfrac{2}{3} \\
  \end{array}
\right)c_{2\,3}\left(
                 \begin{array}{c}
                   -\dfrac{A\a}{2} \\ \\
                   -\dfrac{2A\a}{3}  \\ \\
                   1 \\
                 \end{array}
               \right)=\left(
                         \begin{array}{c}
                           0 \\ \\
                           0 \\ \\
                           0 \\
                         \end{array}
                       \right).
\ee
Eqs.~(\ref{g_B1x})-(\ref{g_B1w}) then imply that as
$\Upsilon\rightarrow 0$, \be a\rightarrow 0, \quad a'\rightarrow
\infty, \quad \rho\rightarrow \infty. \ee This asymptotic behavior
corresponds to a general solution of a flat brane that is valid
around a collapse type I singularity.
We thus 
regain a behavior similar to the one met in Subsection 2.1 for the case
of a flat brane in a scalar field bulk.

\subsection{Collapse type II singularities}
In this Subsection, we show that for a curved brane ($k=\pm 1$) the
long-term (distance) behavior of all solutions which depend on the
asymptotics near finite-distance singularities turn out to be of a
very different nature. In particular, we shall show that the
balances $_{\gamma}\mathcal{B}_{2}$ for $\gamma<-1/2$,
$_{-1/2}\mathcal{B}_{3}$ for $r<-2$ and $_{-1/2}\mathcal{B}_{4}$ (as
we have already mentioned $_{\gamma}\mathcal{B}_{2}$ and
$_{-1/2}\mathcal{B}_{3}$ correspond to a curved brane whereas
$_{-1/2}\mathcal{B}_{4}$ corresponds to a flat or curved brane),
imply the existence of a collapse type IIa, b or c singularity. This
is in sharp contrast to the asymptotic behavior found for a curved
brane in the presence of a bulk scalar field (see Subsection 2.2),
wherein there are no finite-distance singularities.

For the balance $_{\gamma}\mathcal{B}_{2}$ we find that \be
_{\gamma}\mathcal{K}_{2}=D\mathbf{f}\left(\a,\a,0\right)
-\textrm{diag}\left(1,0,-2\right)= \left(
  \begin{array}{ccc}
    -1 & 1 & 0 \\
    0 & 0 & -\dfrac{2}{3}A\a (1+2\gamma)\\
    0 & 0 & -2(1+2\gamma) \\
  \end{array}
\right), \ee and hence, \be
\textrm{spec}(_{\gamma}\mathcal{K}_{2})=\{-1,0,-2(1+2\gamma)\}. \ee
We note that the third arbitrary constant appears at the value
$j=-2(1+2\gamma)$, $\gamma<-1/2$. After substituting the forms, \be
x=\Sigma_{j=0}^{\infty}\,c_{j1}\Upsilon^{j+1},\quad
y=\Sigma_{j=0}^{\infty}\,c_{j2}\Upsilon^{j},\quad
w=\Sigma_{j=0}^{\infty}\,c_{j3}\Upsilon^{j-2}, \ee in the system
(\ref{syst2a})-(\ref{syst2c}), to proceed
we may try giving different values to $\gamma$: Inserting the value
$\gamma =-3/4$ in the system for concreteness we meet a third
arbitrary constant at $j=1$
($\textrm{spec}(_{-3/4}\mathcal{K}_{2})=\{-1,0,1\}$). We then arrive
at the following asymptotic forms of the solution: \bq
\label{-3/4_B2x}
x&=& \a\Upsilon+\frac{A\a}{6}c_{1\,3}\Upsilon^{2}+\cdots,\\
y&=& \a+\frac{A\a}{3} c_{1\,3}\Upsilon+\cdots,\\
\label{-3/4_B2w} w&=&  c_{1\,3}\Upsilon^{-1}+\cdots, \eq where
$c_{1\,3}\neq 0$ \footnote{If we do not set from the beginning
$\gamma=-3/4$ but instead we let $\gamma$ be arbitrary, then in the
last step of the calculations at the $j=1$ level we find that either
$c_{1\,3}=0$ or $\gamma=-3/4$.}. We need to check the validity of
the compatibility condition for $j=1$. But this is trivially
satisfied since \be
(_{-3/4}\mathcal{K}_{2}-\mathcal{I}_{3})\mathbf{c}_{1}= \left(
  \begin{array}{ccc}
    -2 & 1  & 0 \\
     0 & -1 & A\a/3 \\
     0 & 0  & 0 \\
  \end{array}
\right)c_{1\,3}\left(
                 \begin{array}{c}
                   A\a/6 \\
                   A\a/3  \\
                   1 \\
                 \end{array}
               \right)=\left(
                         \begin{array}{c}
                           0 \\
                           0 \\
                           0 \\
                         \end{array}
                       \right).
\ee

The series expansions in eqs.~(\ref{-3/4_B2x})-(\ref{-3/4_B2w}) are
therefore valid and we conclude that as $\Upsilon\rightarrow 0$, \be
\label{-3/4B2} a\rightarrow 0, \quad a'\rightarrow\a, \quad
\rho\rightarrow \infty, \quad \alpha\neq 0. \ee This is a collapse
type IIc singularity. It will follow from the analysis below that
the behavior of $\rho$ depends on our choice of $\gamma$ (thus
giving rise to three possible subcases of a type II singularity).
Indeed, choosing for instance $\gamma=-1$
($\textrm{spec}(_{-1}\mathcal{K}_{2})=\{-1,0,2\}$), we find that the
solution is given by the forms, \bq \label{-1_B2x}
x&=& \a\Upsilon+\frac{A\a}{9}c_{2\,3}\Upsilon^{3}+\cdots,\\
y&=& \a+\frac{A\a}{3} c_{2\,3}\Upsilon^{2}+\cdots,\\
\label{-1_B2w} w&=&  c_{2\,3}+\cdots, \eq where $c_{2\,3}\neq 0$
\footnote{Had we let $\gamma$ be arbitrary we would have found that
in the step $j=2$ of the procedure either $c_{2\,3}=0$ or
$\gamma=-1$.}. Note that the compatibility condition is satisfied
here as well since \be
(_{-1}\mathcal{K}_{2}-2\mathcal{I}_{3})\mathbf{c}_{2}= \left(
  \begin{array}{ccc}
    -3 & 1  & 0 \\
     0 & -2 & 2A\a/3 \\
     0 & 0  & 0 \\
  \end{array}
\right)c_{2\,3}\left(
                 \begin{array}{c}
                   A\a/9 \\
                   A\a/3 \\
                   1 \\
                 \end{array}
               \right)=\left(
                         \begin{array}{c}
                           0 \\
                           0 \\
                           0 \\
                         \end{array}
                       \right).
\ee We see that as $\Upsilon\rightarrow 0$, \be a\rightarrow 0,
\quad a'\rightarrow\a, \quad \rho\rightarrow c_{2\,3}, \quad \a\neq
0. \ee This is a collapse type IIb singularity in our terminology
and is clearly different from (\ref{-3/4B2}).

A yet different behavior is met if we choose for instance
$\gamma=-5/4$. The $\mathcal{K}$-exponents are given by
$\textrm{spec}(_{-5/4}\mathcal{K}_{2})=\{-1,0,3\}$, and the series
expansions become, \bq \label{-5/4_B2x}
x&=& \a\Upsilon+\frac{A\a}{12}c_{3\,3}\Upsilon^{4}+\cdots,\\
y&=& \a+\frac{A\a}{3} c_{3\,3}\Upsilon^{3}+\cdots,\\
\label{-5/4_B2w} w&=&  c_{3\,3}\Upsilon+\cdots,  \eq where
$c_{3\,3}\neq 0$ \footnote{Here again, had we let $\gamma$ be
arbitrary we would have found that in the step $j=3$ of the
procedure, either $c_{3\,3}=0$, or $\gamma=-5/4$.}. These expansions
are valid locally around the singularity since the compatibility
condition holds true because \be
(_{-5/4}\mathcal{K}_{2}-3\mathcal{I}_{3})\mathbf{c}_{1}= \left(
  \begin{array}{ccc}
    -4 & 1  & 0 \\
     0 & -3 & A\a \\
     0 & 0  & 0 \\
  \end{array}
\right)c_{3\,2}\left(
                 \begin{array}{c}
                   A\a/12 \\
                   A\a/3  \\
                   1 \\
                 \end{array}
               \right)=\left(
                         \begin{array}{c}
                           0 \\
                           0 \\
                           0 \\
                         \end{array}
                       \right).
\ee For $\Upsilon\rightarrow 0$, we have that \be a\rightarrow 0,
\quad a'\rightarrow\a, \quad \rho\rightarrow 0, \quad \a\neq 0, \ee
which means that this is a collapse type IIa singularity. This
balance therefore leads to the asymptotic behavior of a particular
solution describing a curved brane approaching a collapse type II
singularity, {\it i.e.} $a\rightarrow 0$ and $a'\rightarrow \alpha$.
The behavior of the density of the perfect fluid varies
dramatically: we can have an infinite density, a constant density,
or even no flow of `energy' at all as we approach the
finite-distance singularity into the extra dimension at $Y_{s}$,
depending on the values of
the $\gamma$ parameter.

We now turn to an analysis of the balances $_{-1/2}\mathcal{B}_{3}$,
for $r<-2$, and $_{-1/2}\mathcal{B}_{4}$. The $\mathcal{K}$-matrix
for $_{-1/2}\mathcal{B}_{3}$ is \be
_{-1/2}\mathcal{K}_{3}=D\mathbf{f}\left(\a,\a,0\right)
-\textrm{diag}(1,0,r)= \left(
  \begin{array}{ccc}
    -1 & 1 & 0 \\
    0 & 0 & 0\\
    0 & 0 & -2-r \\
  \end{array}
\right), \ee and hence, \be
\textrm{spec}(_{-1/2}\mathcal{K}_{3})=\{-1,0,-2-r\}. \ee Taking
$-2-r>0$, we have two non-negative $\mathcal{K}$-exponents. (The
case $-2-r<0$ is considered later, in Subsection 3.5, since it is quite
different, and it does not imply the existence of a finite-distance
singularity.) For $r=-3$ as an example, we substitute the forms \be
x=\Sigma_{j=0}^{\infty}\,c_{j1}\Upsilon^{j+1},\quad
y=\Sigma_{j=0}^{\infty}\,c_{j2}\Upsilon^{j},\quad
w=\Sigma_{j=0}^{\infty}\,c_{j3}\Upsilon^{j-3}, \ee and arrive at the
expansions \bq \label{-1/2_B_3x}
x&=& \a\Upsilon+\cdots,\\
y&=& \a+\cdots,\\
\label{-1/2_B_3w} w&=&  c_{1\,3}\Upsilon^{-2}+\cdots. \eq The
compatibility condition is satisfied because \be
(_{-1/2}\mathcal{K}_{3}-\mathcal{I}_{3})\mathbf{c}_{1}= \left(
  \begin{array}{ccc}
    -2 & 1  & 0 \\
     0 & -1 & 0 \\
     0 & 0  & 0 \\
  \end{array}
\right)c_{1\,3}\left(
                 \begin{array}{c}
                   0 \\
                   0  \\
                   1 \\
                 \end{array}
               \right)=\left(
                         \begin{array}{c}
                           0 \\
                           0 \\
                           0 \\
                         \end{array}
                       \right),
\ee and so the expansions (\ref{-1/2_B_3x})-(\ref{-1/2_B_3w}) are
valid ones in the vicinity of the singularity. 
The general behavior of the solution is then characterized by the
asymptotic forms \be a\rightarrow 0, \quad a'\rightarrow \a,
\quad\rho \rightarrow\infty, \quad \a\neq 0. \ee The balance
$_{-1/2}\mathcal{B}_{3}$ for $r<-2$ implies therefore the existence
of a collapse type IIc singularity during the dynamical evolution of
the curved brane living (and moving) in this specific perfect fluid
bulk.

The balance $_{-1/2}\mathcal{B}_{4}$, on the other hand, is one with
\be _{-1/2}\mathcal{K}_{4}=D\mathbf{f}\left(\a,\a,\d\right)
-\textrm{diag}(1,0,-2)= \left(
  \begin{array}{ccc}
    -1 & 1 & 0 \\
    0 & 0 & 0\\
    \dfrac{2\d}{\a} & -\dfrac{2\d}{\a} & 0 \\
  \end{array}
\right), \ee and \be
\textrm{spec}(_{-1/2}\mathcal{K}_{4})=\{-1,0,0\}. \ee We note that
the double multiplicity of the zero eigenvalue reflects the fact
that there were already two arbitrary constants, $\a$ and $\d$ in
this balance (recall though that $\d$ had to be sacrificed in order
for this balance to satisfy the constraint (\ref{constraint3})). We
can thus write \bq \label{-1/2_B4x}
x&=& \a\Upsilon+\cdots,\\
y&=& \a+\cdots,\\
\label{-1/2_B4w} w&=&  \d\Upsilon^{-2}+\cdots, \eq so that as
$\Upsilon\rightarrow 0$, a collapse type IIc singularity develops,
{\it i.e.}, \be a\rightarrow 0,\quad a'\rightarrow \a,
\quad\rho\rightarrow\infty, \quad \alpha\neq 0. \ee
\subsection{Big rip singularities}
In this Subsection we return to the study of the balance $_{\gamma}\mathcal{B}_{1}$
but focus on different $\gamma$ values. In particular, we show that
when $\gamma<-1$, a flat brane develops a big rip singularity at a
finite distance. This new asymptotic behavior implied by the balance
$_{\gamma}\mathcal{B}_{1}$ (when $\gamma<-1$) is equally general to
the one found in Subsection 3.1.

For purposes of illustration, let us take $\gamma=-2$. Then the
balance $_{-2}\mathcal{B}_{1}$ and the
$_{-2}\mathcal{K}_{1}$-exponents read, respectively, \bq
_{-2}\mathcal{B}_{1}&=&\{(\a,-\a/2,3/(8A)),(-1/2,-3/2,-2)\},\\
\textrm{spec}(_{-2}\mathcal{K}_{1})&=&\{-1,0,3\}. \eq Substituting
the value $\gamma=-2$ in our basic system given by
eqs.~(\ref{syst2a})-(\ref{syst2c}), and also the forms \be
x=\Sigma_{j=0}^{\infty}\,c_{j1}\Upsilon^{j-1/2},\quad
y=\Sigma_{j=0}^{\infty}\,c_{j2}\Upsilon^{j-3/2},\quad
w=\Sigma_{j=0}^{\infty}\,c_{j3}\Upsilon^{j-2}, \ee we expect to meet
the third arbitrary constant at $j=3$. Indeed we find: \bq
\label{ph_B1x}
x&=& \a\Upsilon^{-1/2}+\frac{2}{3}A\a c_{3\,3}\Upsilon^{5/2}+\cdots,\\
y&=& -\frac{\a}{2}\Upsilon^{-3/2}+\frac{5}{3}A \a c_{3\,3}\Upsilon^{3/2}+\cdots,\\
\label{ph_B1w} w&=&
\frac{3}{8A}\Upsilon^{-2}+c_{3\,3}\Upsilon+\cdots, \quad
c_{3\,3}\neq 0. \eq The compatibility condition is trivially
satisfied for $j=3$, since the product
$(_{-2}\mathcal{K}_{1}-3\mathcal{I}_{3})\mathbf{c}_{3}$ is
identically zero: \be
(_{-2}\mathcal{K}_{1}-3\mathcal{I}_{3})\mathbf{c}_{3}= \left(
  \begin{array}{ccc}
    -\dfrac{5}{2} & 1 & 0 \\ \\
    \dfrac{3}{4} & -\dfrac{3}{2} & 2A\a \\ \\
    \dfrac{3}{4A\a} & \dfrac{3}{2A\a} & -3 \\
  \end{array}
\right)c_{3\,3}\left(
                 \begin{array}{c}
                   \dfrac{2}{3}A\a \\ \\
                   \dfrac{5}{3}A\a  \\ \\
                   1 \\
                 \end{array}
               \right)=\left(
                         \begin{array}{c}
                           0 \\
                           0 \\
                           0 \\
                         \end{array}
                       \right).
\ee
The series expansions given by eqs.~(\ref{ph_B1x})-(\ref{ph_B1w})
are therefore valid asymptotically for $\Upsilon\rightarrow 0$ so
that we end up with the asymptotic forms \be a\rightarrow
\infty,\quad a'\rightarrow -\infty, \quad\rho\rightarrow \infty. \ee
We therefore conclude that the balance $_{\gamma}\mathcal{B}_{1}$ 
leads to a general solution in which a flat brane develops a big rip
singularity after `traveling' for a finite distance when the bulk
perfect fluid satisfies a phantom-like equation of state, {\it i.e.}
$\gamma<-1$. Note that using the analogy between the warp factor of
our braneworld and the scale factor of an expanding universe, we can
say that this singularity bares many similarities to the one studied
in Refs.~\cite{cald99},~\cite{cald03},~\cite{go03}, since it is also
characterized by all quantities $a$, $a'$, $\rho$, and consequently
$P$, becoming asymptotically divergent. Thus, the results in this
Subsection indicate that a flat brane traveling in a $\gamma<-1$ fluid
bulk develops a big rip singularity. This implements the behavior
found in Subsection 3.1 of the present paper, wherein the same brane
moving in a $\gamma>-1/2$ fluid bulk `disappears' in a big bang-type
singularity.
\subsection{Sudden behavior}
As our penultimate mode of approach to the finite-distance
singularity, we examine here the balance
$_{-1/2}\mathcal{B}_{5}=\{(\a,0,0), (0,-1,r)\}$. This balance has
\be _{-1/2}\mathcal{K}_{5}=D\mathbf{f}\left(\a,0,0\right)
-\textrm{diag}(0,-1,r)= \left(
  \begin{array}{ccc}
    0 & 1 & 0 \\
    0 & 1 & 0\\
    0 & 0 & -r \\
  \end{array}
\right), \ee and \be
\textrm{spec}(_{-1/2}\mathcal{K}_{5})=\{1,0,-r\}, \ee so we shall
have to set $r=1$ in order to have the necessary $-1$ eigenvalue
corresponding to the arbitrary position of the `singularity',
$Y_{s}$. After substitution of the forms \be
x=\Sigma_{j=0}^{\infty}\,c_{j1}\Upsilon^{j},\quad
y=\Sigma_{j=0}^{\infty}\,c_{j2}\Upsilon^{j-1},\quad
w=\Sigma_{j=0}^{\infty}\,c_{j3}\Upsilon^{j+1}, \ee we find that the
solution reads \bq \label{-1/2_B4xn}
x&=& \a+c_{1\,1}\Upsilon+\cdots,\\
y&=& c_{1\,1}+\cdots,\\
\label{-1/2_B4wn} w&=&  0+\cdots. \eq The compatibility condition is
satisfied since \be
(_{-1/2}\mathcal{K}_{5}-\mathcal{I}_{3})\mathbf{c}_{1}= \left(
  \begin{array}{ccc}
    -1 & 1  & 0 \\
     0 & 0  & 0 \\
     0 & 0  & 0 \\
  \end{array}
\right)c_{1\,1}\left(
                 \begin{array}{c}
                   1 \\
                   1 \\
                   0 \\
                 \end{array}
               \right)=\left(
                         \begin{array}{c}
                           0 \\
                           0 \\
                           0 \\
                         \end{array}
                       \right),
\ee and we see that as $\Upsilon\rightarrow 0$, \be a\rightarrow \a,
\quad a'\rightarrow c_{1\,1}, \quad \rho\rightarrow 0, \quad \alpha\neq
0. \ee This clearly indicates that the brane experiences the
so-called sudden behavior (cf.~\cite{ba04}).
\subsection{Behavior at infinity}
A qualitatively  different picture than what we have already encountered
in our analysis of brane singularities in a fluid bulk is
attained by exploiting either the balance $_{\gamma}\mathcal{B}_{1}$ with
$-1<\gamma<-1/2$, or the balance $_{\gamma}\mathcal{B}_{2}$ with
$\gamma >-1/2$, or the balance $_{-1/2}\mathcal{B}_{3}$ with $r>-2$.
We show in this Subsection that these three balances \emph{and only these}
offer the possibility of avoiding the finite-distance singularities
met before and may describe the behavior of our model at infinity.

We begin with the balance $_{\gamma}\mathcal{B}_{1}$ when
$-1<\gamma<-1/2$. Choosing for instance $\gamma =-4/5$, we find
$\textrm{spec}(_{-4/5}\mathcal{K}_{1})=\{-1,0,-3\}$, and hence we
may expand $(x,y,w)$ in descending powers in order to meet the
arbitrary constants appearing at $j=-1$ and $j=-3$, {\it i.e.} \be
x=\Sigma_{j=0}^{-\infty}\,c_{j1}\Upsilon^{j+5/2},\quad
y=\Sigma_{j=0}^{-\infty}\,c_{j2}\Upsilon^{j+3/2},\quad
w=\Sigma_{j=0}^{-\infty}\,c_{j3}\Upsilon^{j-2}. \ee We find: \bq
\label{-4/5_B1x} x&=& \a\Upsilon^{5/2}+c_{-1\,1}\Upsilon^{3/2}
+3/(10\a)c_{-1\,1}^{2}\Upsilon^{1/2}+c_{-3\,1}\Upsilon^{-1/2}+\cdots,\\
y&=& 5\a/2\Upsilon^{3/2}+3/2 c_{-1\,1}\Upsilon^{1/2}
+3/(20\a)c_{-1\,1}^{2}\Upsilon^{-1/2}
-1/2c_{-3\,1}\Upsilon^{-3/2}+\cdots,\\
\label{-4/5_B1w} \nonumber w&=&
75/(8A)\Upsilon^{-2}-15/(2A\a)c_{-1\,1}\Upsilon^{-3}+
9/(2A\a^{2})c_{-1\,1}^{2}\Upsilon^{-4}+\\
&\,\,\,&+\left(-15/(2A\a)c_{-3\,1}-9/(4A\a^{3})c_{-1\,1}^{3}\right)\Upsilon^{-5}
+\cdots. \eq The compatibility conditions at $j=-1$ is satisfied
since \be
(_{-4/5}\mathcal{K}_{1}+\mathcal{I}_{3})\mathbf{c}_{-1}=\left(
  \begin{array}{ccc}
     -3/2      & 1           & 0 \\
     15/4      & -1/2        & 2A\a/5 \\
     75/(4A\a) & -15/(2A\a)  & 1 \\
  \end{array}
\right)c_{-1\,1}\left(
                 \begin{array}{c}
                   1\\
                   3/2  \\
                   -15/(2A\a) \\
                 \end{array}
               \right)=\left(
                         \begin{array}{c}
                           0 \\
                           0 \\
                           0 \\
                         \end{array}
                       \right).
\ee But for $j=-3$ we find \bq \nonumber
&&(_{-4/5}\mathcal{K}_{1}+3\mathcal{I}_{3})\mathbf{c}_{-3}=\\
\nonumber &=&\left(
  \begin{array}{ccc}
    1/2       & 1          & 0 \\
    15/4      & 3/2        & 2A\a/5 \\
    75/(4A\a) & -15/(2A\a) & 3 \\
  \end{array}
\right)\left(
                 \begin{array}{c}
                       c_{-3\,1}\\
                   -1/2c_{-3\,1} \\
                   -15/(2A\a)c_{-3\,1}-9/(4A\a^{3})c_{-1\,1}^{3} \\
                 \end{array}
               \right)\\
               &=&\left(
                         \begin{array}{c}
                           0 \\
                           -9/(10\a^{2})c_{-1\,1}^{3} \\
                           -27/(4A\a^{3})c_{-1\,1}^{3} \\
                         \end{array}
                       \right)=P_{-3}.
\eq An eigenvector corresponding to the eigenvalue $j=-3$ is
$v^{\top}=(-2A\a/15,A\a/15,1)$, and hence we have \be v^{\top}\cdot
P_{-3}\neq 0, \ee unless $c_{-1\,1}=0$. In order to satisfy the
compatibility condition at $j=-3$ we set $c_{-1\,1}=0$. The solution
(\ref{-4/5_B1x})-(\ref{-4/5_B1w}) with $c_{-1\,1}=0$ reads \bq
\label{-4/5_B1x2}
x&=& \a\Upsilon^{5/2}+c_{-3\,1}\Upsilon^{-1/2}+\cdots,\\
y&=& 5\a/2\Upsilon^{3/2}-1/2c_{-3\,1}\Upsilon^{-3/2}+\cdots,\\
\label{-4/5_B1w2}
w&=&75/(8A)\Upsilon^{-2}-15/(2A\a)c_{-3\,1}\Upsilon^{-5}+\cdots \eq
and it is a particular solution containing two arbitrary constants.
As $S\equiv 1/\Upsilon\rightarrow\infty$, we conclude that \be
a\rightarrow \infty, \quad a'\rightarrow\infty, \quad
\rho\rightarrow \infty, \ee and we can therefore avoid the
finite-distance singularity in this case.

Next we examine the balance $_{\gamma}\mathcal{B}_{2}$ when
$\gamma>-1/2$. For $\gamma=0$, we have that
$\textrm{spec}(_{0}\mathcal{K}_{2})=\{-1,0,-2\}$, and hence we
substitute \be x=\Sigma_{j=0}^{-\infty}\,c_{j1}\Upsilon^{j+1},\quad
y=\Sigma_{j=0}^{-\infty}\,c_{j2}\Upsilon^{j},\quad
w=\Sigma_{j=0}^{-\infty}\,c_{j3}\Upsilon^{j-2}, \ee and find: \bq
\label{0_B2x}
x&=& \a\Upsilon+c_{-1\,1}-A\a/3c_{-2\,3}\Upsilon^{-1}+\cdots,\\
y&=& \a+A\a/3 c_{-2\,3}\Upsilon^{-2}+\cdots,\\
\label{0_B2w} w&=&  c_{-2\,3}\Upsilon^{-4}+\cdots. \eq The
compatibility conditions at $j=-1$ and $j=-2$ are indeed satisfied
since \be (_{0}\mathcal{K}_{2}+\mathcal{I}_{3})\mathbf{c}_{-1}=
\left(
  \begin{array}{ccc}
     0 & 1  & 0 \\
     0 & 1  & -2A\a/3 \\
     0 & 0  & -1 \\
  \end{array}
\right)c_{-1\,1}\left(
                 \begin{array}{c}
                   1\\
                   0  \\
                   0 \\
                 \end{array}
               \right)=\left(
                         \begin{array}{c}
                           0 \\
                           0 \\
                           0 \\
                         \end{array}
                       \right),
\ee and \be (_{0}\mathcal{K}_{2}+2\mathcal{I}_{3})\mathbf{c}_{-2}=
\left(
  \begin{array}{ccc}
    1 & 1  & 0 \\
     0 & 2 & -2A\a/3 \\
     0 & 0  & 0 \\
  \end{array}
\right)c_{-2\,3}\left(
                 \begin{array}{c}
                   -A\a /3\\
                   A\a/3  \\
                   1 \\
                 \end{array}
               \right)=\left(
                         \begin{array}{c}
                           0 \\
                           0 \\
                           0 \\
                         \end{array}
                       \right).
\ee As $S\equiv 1/\Upsilon\rightarrow\infty$, we conclude that \be
a\rightarrow \infty, \quad a'\rightarrow\infty, \quad
\rho\rightarrow \infty, \ee and the finite-distance singularity in
shifted at an infinite distance.

We now move on to the balance $_{-1/2}\mathcal{B}_{3}$, $r>-2$. In
this case we have two negative $\mathcal{K}$-exponents. If
we choose the value $r=0$, then the spectrum is found to be \be
\textrm{spec}(_{-1/2}\mathcal{K}_{3})=\{-1,0,-2\}, \ee and so
inserting the forms \be
x=\Sigma_{j=0}^{-\infty}\,c_{j1}\Upsilon^{j+1},\quad
y=\Sigma_{j=0}^{-\infty}\,c_{j2}\Upsilon^{j},\quad
w=\Sigma_{j=0}^{-\infty}\,c_{j3}\Upsilon^{j}, \ee we obtain \bq
\label{-1/2_B3x}
x&=& \a\Upsilon+c_{-1\,1},\\
y&=& \a,\\
\label{-1/2_B3w} w&=&  c_{-2\,3}\Upsilon^{-2}+\cdots, \eq which
validates the compatibility conditions at $j=-1$ and $j=-2$ since
\be (_{-1/2}\mathcal{K}_{3}+\mathcal{I}_{3})\mathbf{c}_{-1}= \left(
  \begin{array}{ccc}
      0 & 1  &  0 \\
      0 & 1  &  0 \\
      0 & 0  & -1 \\
  \end{array}
\right)c_{-1\,1}\left(
                 \begin{array}{c}
                   1 \\
                   0 \\
                   0 \\
                 \end{array}
               \right)=\left(
                         \begin{array}{c}
                           0 \\
                           0 \\
                           0 \\
                         \end{array}
                       \right),
\ee and \be
(_{-1/2}\mathcal{K}_{3}+2\mathcal{I}_{3})\mathbf{c}_{-2}= \left(
  \begin{array}{ccc}
    1 & 1  & 0 \\
     0 & 2  & 0 \\
     0 & 0  & 0 \\
  \end{array}
\right)c_{-2\,3}\left(
                 \begin{array}{c}
                   0 \\
                   0 \\
                   1 \\
                 \end{array}
               \right)=\left(
                         \begin{array}{c}
                           0 \\
                           0 \\
                           0 \\
                         \end{array}
                       \right).
\ee We see that as $S\equiv 1/\Upsilon\rightarrow\infty$, \be
a\rightarrow c_{-1\,1}, \quad a'\rightarrow\a, \quad \rho\rightarrow
\infty, \quad \a\neq 0, \ee so that the balance
$_{-1/2}\mathcal{B}_{3}$ for $r>-2$ also offers the possibility of
escaping the finite-distance singularities. Hence in such cases we
find a regular (singularity free) evolution of the brane as it
travels in the bulk filled with the type of matter considered above.
\section{Conclusions}
In this paper we studied a braneworld consisting of a 3-brane
embedded in a five-dimensional bulk space, filled either with a
scalar field or with an analogue of  perfect fluid, giving special
emphasis in the possible formation of finite-distance singularities
away from the brane into the bulk.

For the case of a scalar field in the bulk, we have shown that the dynamical behavior
of the model strongly depends on the spatial geometry of the brane, in particular
whether it is flat or not. For a flat brane the model experiences a
finite-distance singularity, that we call a collapse type I
singularity ($a\rightarrow 0$, $a'\rightarrow \infty$, $\phi'\rightarrow\infty$,
as $Y\rightarrow Y_{s}$), toward which all the vacuum energy decays, whereas for
a curved brane the model avoids the singularity which is now located
at an infinite distance.
Note that in this case, the collapse type I singularity 
is the only 
possible one that can develop at a finite distance from the (flat) brane.

In the second part of this paper, we studied the dynamical
`evolution' of a braneworld where we replaced the scalar field with
a `perfect fluid' possessing a general equation of state
$P=\gamma\rho$, characterized by the constant parameter $\gamma$.
For a flat brane, we find that it is possible to have within finite
distance from the brane a collapse type I singularity met previously
in the case of a scalar field (where this singularity was the
\emph{only} type possible, as we already mentioned above). In the
fluid case, we showed that in addition to that singularity which
appears inevitably in all flat brane solutions with $\gamma>-1/2$,
there are two other new types (for a flat brane): The first one is
the very distinct big rip singularity which occurs with
$a\rightarrow\infty$, $a'\rightarrow -\infty$,
$\rho\rightarrow\infty$, only when a phantom type equation of state
with $\gamma<-1$ is considered. The second one is a collapse type
IIc singularity which may be described by the behavior $a\rightarrow
0$, $a'\rightarrow \alpha$ and $\rho\rightarrow\infty$. This is less
general than the collapse type I and the big rip singularities and
it arises only when $\gamma=-1/2$. Besides these singular solutions,
we found the surprising result of flat branes without
finite-distance singularities in the region $-1<\gamma\le -1/2$.
Moreover, for $\gamma=-1/2$, there is also a solution with sudden
behavior having $a$ and $a'$ finite and vanishing density $\rho\to
0$~\cite{ba04}.

In contrast to the bulk scalar field case where all curved brane
solutions were regular, in the case of a perfect fluid in the bulk we
found also singular such solutions.
The possible corresponding
finite-distance singularities are the ones comprising the collapse
type II class. These are singularities with $a\rightarrow 0$,
$a'\rightarrow \alpha$ and $\rho\rightarrow 0, \rho_{s},\infty$
(corresponding to types IIa, b and c, respectively).
The interesting feature of this class of singularities is that it
allows the `energy' leak into the extra dimension to vary and be
monitored each time by the $\gamma$ parameter that defines the type
of fluid; they all arise in the region $\gamma\le -1/2$. On the
other hand, we showed that for a curved brane the possibility of
avoiding the finite-distance singularities that was offered in
the scalar field case is still valid here, but only in the region $\gamma\ge
-1/2$.

For illustration, we present a summary of all different behaviors we
found for flat and curved branes in the table below, using the
notation for the various singularities introduced in Section 2 after Eq.~(\ref{constraint1})
and the balances (\ref{preveq})-(\ref{Bfive}).
\begin{center}
\begin{tabular}{|c|c|c|c|c|}
\hline
equation of state &\mco{2}{c|}{flat brane} & \mco{2}{c|}{curved brane} \\
\hline $P=\gamma\rho$ & type & balance & type & balance \\ \hline
$\gamma>-1/2$ & singular type I & $_{\gamma}\mathcal{B}_{1}$ & {\em
regular} & $_{\gamma}\mathcal{B}_{2}$ at $\infty$ \\ \hline
$\gamma=-1/2$ & singular IIc & $_{-1/2}\mathcal{B}_{4}$ & {\em
regular} &
$_{-1/2}\mathcal{B}_{3}\,$, $r>-2$ at $\infty$ \\
 & {\em regular} sudden & $_{-1/2}\mathcal{B}_{5}$ & singular IIc &
 $_{-1/2}\mathcal{B}_{3\,}$, $r<-2$ or $_{-1/2}\mathcal{B}_{4}$ \\ \hline
$-1<\gamma<-1/2$ & {\em regular} & $_{\gamma}\mathcal{B}_{1}$ at
$\infty$ & singular IIa,b,c & $_{\gamma}\mathcal{B}_{2}$ \\ \hline
$\gamma=-1$ & no solution & & singular IIb &
$_{\gamma}\mathcal{B}_{2}$ \\ \hline $\gamma<-1$ & singular big rip
& $_{\gamma}\mathcal{B}_{1}$ & singular IIa &
$_{\gamma}\mathcal{B}_{2}$ \\
\hline
\end{tabular}

\end{center}

An open question is whether there exist any physical constraints on
$\gamma$ analogous to the weak and strong energy conditions of
matter in ordinary perfect fluid cosmology. A related question is to
find possible field theory realizations of the `exotic' regions of
$\gamma\le -1/2$, where interesting solutions with unexpected
behavior were found. The most important issue of course is to
clarify the possibility of singularity avoidance at finite distance
in flat brane solutions. There is no reason why  the non-singular
behavior for flat branes discovered here  should not persist for
arbitrary values of the brane tension and, indeed, it is to be
expected that only particular asymptotic modes of behavior, {\it
i.e.}, specific detailed forms of asymptotic solutions, would depend
on such values. Thus, the self-tuning mechanism appears to be a
property of a general (non-singular) flat brane solution that
depends on two arbitrary constants in the region $-1<\gamma<-1/2$
(three for the general solution with sudden behavior when
$\gamma=-1/2$). Similarly, as we have shown here, the existence of
singular curved brane solutions in some regions of $\gamma$ is
independent of the sign of the scalar curvature (as long as the
latter remains nonzero for curved branes), but the particular way of
asymptotic approach to the singularity is sensitive to that sign and
it may therefore change with different values of the brane tension.

It would also be interesting to further investigate whether the
properties of finite-distance singularities (and their possible
avoidance) encountered here continue to emerge in more general
systems, such as the case in which a scalar field \emph{coexists}
with a perfect fluid in the bulk~\cite{ack3}. The analysis of this
more involved case that allows for fluid interactions may also shed light to the factors that control
how these two bulk matter components compete on approach to the
singularity, or even predict new types of singularities that might
then become feasible, as well as possible situations where they can
be avoided.

\section*{Acknowledgements}
S.C. and I.K. are grateful to CERN-Theory Division, where part of
their work was done, for making their visits there possible and for
allowing them to use its excellent facilities. The work of I.A. was
supported in part by the European Commission under the  ERC Advanced
Grant 226371 and the contract PITN-GA-2009-237920 and in part by the
CNRS grant GRC APIC PICS 3747.

\appendix
\section{Appendix: The method of asymptotic splittings}
We refer briefly here to the basic steps of the method of asymptotic
splittings. A detailed analysis can be found in Ref.~\cite{skot}.

Consider a system of $n$ first order ordinary differential equations
\be \label{arb_syst} \mathbf{x'}=\mathbf{f}(\mathbf{x}), \ee where
$\mathbf{x}=(x_{1},\ldots,x_{n})\in \mathbb{R}^{n}$,
$\mathbf{f}(\mathbf{x})=(f_{1}(\mathbf{x}),\ldots,f_{n}(\mathbf{x}))$
and $'\equiv\frac{d}{dY}$, $Y$ being the independent variable. In
this paper, we refrain from calling $Y$ a time variable and giving
it the interpretation of time. Since we are interested in
singularities located at a \emph{distance} from the brane and into
the bulk, it seems more appropriate to talk about finite-distance
singularities and give to the $Y$ variable a spatial interpretation.
The general solution of the above system contains $n$ arbitrary
constants and describes all possible behaviors of the system
starting from arbitrary initial data. Any particular solution of
(\ref{arb_syst}), on the other hand, contains less than $n$
arbitrary constants and describes a possible behavior of the system
emerging from a proper subset of initial data space.

We say that a solution of the dynamical system (\ref{arb_syst})
exhibits a finite-distance singularity if there exists a $Y_{s}\in
\mathbb{R}$ and a $\mathbf{x}_{0}\in \mathbb{R}^{n}$ such that
\be \lim_{Y\rightarrow
Y_{s}}\|\mathbf{x}(Y;\mathbf{x}_{0})\|\rightarrow\infty,
\ee where $\|\centerdot\|$ is any $L^{p}$ norm. The purpose of
singularity analysis (cf.~\cite{skot},~\cite{goriely}) 
is to build series expansions of solutions around the presumed
position of a singularity at $Y_{s}$ in order to study the different
asymptotic behaviors of the solutions of the system
(\ref{arb_syst}) as one approaches this singularity. In particular,
we look for series expansions of solutions that take the form of a
Puiseux series (any $\log$ terms absent), namely, a series of the
form \be \label{Puiseux_app}
\mathbf{x}=\Upsilon^{\mathbf{p}}\left(\mathbf{a}
+\Sigma_{i=1}^{\infty}\,\mathbf{c}_{i}\Upsilon^{i/s}\right), \ee where
$\Upsilon=Y-Y_{s}$, $\mathbf{p}\in \mathbb{Q}^{n}$,
$s\in\mathbb{N}$.

The method of asymptotic splittings for any system of the form
(\ref{arb_syst}) is realized by taking the following steps:

$\bullet$ First, we find all the possible \emph{weight-homogeneous
decompositions} of the vector field $\mathbf{f}$ by splitting it
into components $\mathbf{f}^{(j)}$: \be
\mathbf{f}=\mathbf{f}^{(0)}+\mathbf{f}^{(1)}+\ldots
+\mathbf{f}^{(k)}, \ee with each of these components being
\emph{weight homogeneous}, that is to say \be
\mathbf{f^{(j)}}(\mathbf{a}\Upsilon^{\mathbf{p}})=\tau^{\mathbf{p}
+\mathbf{1}(q^{(j)}-1)}\mathbf{f}^{(j)}(\mathbf{a}) \quad
j=0,\ldots,k, \ee where $\mathbf{a}\in\mathbb{R}^{n}$ and $q^{(j)}$
are the positive non-dominant exponents that are defined by
(\ref{sub_exp}) below.

$\bullet$ We substitute the forms
$\mathbf{x}=\mathbf{a}\mathbf{\Upsilon}^{\mathbf{p}}$ in the system
$\mathbf{x}'=\mathbf{f}^{(0)}(\mathbf{x})$ in order to find all
possible \emph{dominant balances}, {\it i.e.} finite sets of the
form $\{\mathbf{a},\mathbf{p}\}$. The \emph{order} of each balance
is defined as the number of the nonzero components of $\mathbf{a}$.

$\bullet$ For each of these balances we check the validity of the
following \emph{dominance condition}:
\be \label{dominance}
\lim_{\Upsilon\rightarrow 0}
\frac{\Sigma_{j=1}^{k}\mathbf{f}^{(j)}(\mathbf{a}\Upsilon^{\mathbf{p}})}
{\Upsilon^{\mathbf{p}-1}}=0, \ee and define the non-dominant
exponents $q^{(j)}$, $j=1,\ldots, k$ by the requirement that \be
\label{sub_exp}
\frac{\Sigma_{j=1}^{k}\mathbf{f}^{(j)}(\Upsilon^{\mathbf{p}})}
{\Upsilon^{\mathbf{p}-1}}\sim\Upsilon^{q^{(j)}}. \ee The balances
that cannot satisfy the condition (\ref{dominance}) are then
discarded.

$\bullet$ We compute the Kovalevskaya matrix $\mathcal{K}$ defined
by \be \mathcal{K}=D\mathbf{f}^{(0)}(\mathbf{a})-\textrm{diag}\, \mathbf{p},
\ee where $D\mathbf{f}^{(0)}(\mathbf{a})$ is the Jacobian matrix of
$\mathbf{f}^{(0)}$ evaluated at $\mathbf{a}$.

$\bullet$ We calculate the spectrum of the $\mathcal{K}$-matrix,
$spec(\mathcal{K})$, that is the set of its $n$ eigenvalues also
called the \emph{$\mathcal{K}$-exponents}. The arbitrary constants
of any particular or general solution first appear in those terms in
the series (\ref{Puiseux_app}) whose coefficients $\mathbf{c}_{k}$ have
indices $k=\varrho s$, where $\varrho$ is a non-negative
$\mathcal{K}$-exponent and $s$ is the least common multiple of the
denominators of the set consisting of the non-dominant exponents
$q^{(j)}$ and of the positive $\mathcal{K}$-exponents (cf.~\cite{skot},~\cite{goriely}).
The number of non-negative $\mathcal{K}$-exponents equals therefore
the number of arbitrary constants that appear in the series
expansions of (\ref{Puiseux_app}). There is always the $-1$ exponent
that corresponds to the position of the singularity, $Y_{s}$. A
dominant balance corresponds thus to a general solution if it
possesses $n-1$ non-negative $\mathcal{K}$-exponents (the $n$th
arbitrary constant is the position of the singularity, $Y_{s}$).

$\bullet$ We substitute the Puiseux series: \be
x_{i}=\Sigma_{j=0}^{\infty}\,c_{ji}\Upsilon^{p_{i}+j/s}, i=1,\ldots,
n, \ee in the system (\ref{arb_syst}).

$\bullet$ We find the coefficients $\mathbf{c}_{j}$ by solving the
recursion relations \be
\mathcal{K}\mathbf{c}_{j}-\frac{j}{s}\,\mathbf{c}_{j}=
P_{j}(\mathbf{c}_{1},\ldots, \mathbf{c}_{j-1}) \ee where $P_{j}$ are
polynomials that are read off from the original system.

$\bullet$ We verify that for every $j=\varrho s$, with $\varrho$ a
positive $\mathcal{K}$-exponent, the following compatibility
conditions hold: \be \label{comp_cond} \mathbf{v}^{\top}\cdot
P_{j}=0, \ee where $\mathbf{v}$ is an eigenvector associated
with the positive $\mathcal{K}$-exponent $\varrho$.

$\bullet$ We repeat the procedure for each possible decomposition.

We note that if the compatibility condition above (Eq.~(\ref{comp_cond}))
is violated at some eigenvalue in the
$\textrm{spec}(\mathcal{K})$, then the original Puiseux series
representation of the solution cannot be admitted and instead we
have to use a \emph{$\psi$-series} for each one of the eigenvalues
with this property. This is a series that includes $\log$ terms of
the form \be \mathbf{x}=\Upsilon^{\mathbf{p}}\left(\mathbf{a}
+\Sigma_{i,j=1}^{\infty}\mathbf{c}_{ij}\Upsilon^{i/s}(\Upsilon^{\varrho}\log\Upsilon)^{j/s}\right),
\ee where $\varrho$ is the $\mathcal{K}$-exponent for which the
compatibility condition is violated. The rest of the procedure in
this case is the same as before.

\end{document}